\documentclass[12pt,draftclsnofoot, onecolumn]{IEEEtran}

\usepackage[cmex10]{amsmath}
\usepackage{amsfonts,amssymb,cite,graphicx,delarray,color}
\usepackage{multirow}
\usepackage{algorithmicx}
\usepackage{algorithm}
\usepackage{algpseudocode}
\usepackage{color}
\IEEEoverridecommandlockouts

\newtheorem{theorem}{Theorem}
\newtheorem{corollary}{Corollary}
\newtheorem{remark}{Remark}
\newtheorem{lemma}{Lemma}

\begin{document}

\title{Success Probability and Area Spectral Efficiency in Multiuser MIMO HetNets}
\date{ }
\author{
    \IEEEauthorblockN{Chang~Li,~\IEEEmembership{Student Member,~IEEE}, Jun~Zhang,~\IEEEmembership{Senior Member,~IEEE}, Jeffrey~G.~Andrews~\IEEEmembership{Fellow,~IEEE}, and Khaled~B.~Letaief,~\IEEEmembership{Fellow,~IEEE}}
    \thanks{C. Li, J. Zhang and K. B. Letaief are with the Dept. of ECE at the Hong Kong University of Science and Technology, Hong Kong (email: \{changli, eejzhang, eekhaled\}@ust.hk). K. B. Letaief is also with Hamad bin Khalifa University, Qatar (email: kletaief@hbku.edu.qa). J. G. Andrews is with Wireless Networking and Communications Group (WNCG), the University of Texas at Austin, TX, USA (email: jandrews@ece.utexas.edu).}
    \thanks{This work was supported by the Hong Kong Research Grant Council under Grant No. 610113. This work was presented in part at IEEE International Conference on Communications, London, UK, June 2015 \cite{Li15_ICC}.}
}

\begin{titlepage}
\maketitle
\thispagestyle{empty}
\vspace{-1.5cm}
\begin{abstract}
    We derive a general and closed-form result for the success probability in downlink multiple-antenna (MIMO) heterogeneous cellular networks (HetNets), utilizing a novel Toeplitz matrix representation.  This main result, which is equivalently the signal-to-interference ratio (SIR) distribution, includes multiuser MIMO, single-user MIMO and per-tier biasing for $K$ different tiers of randomly placed base stations (BSs), assuming zero-forcing precoding and perfect channel state information.  The large SIR limit of this result admits a simple closed form that is accurate at moderate SIRs, e.g., above 5 dB.   These results reveal that the SIR-invariance property of SISO HetNets does not hold for MIMO HetNets; instead the success probability may decrease as the network density increases.  We prove that the maximum success probability is achieved by activating only one tier of BSs, while the maximum area spectral efficiency (ASE) is achieved by activating all the BSs.   This reveals a unique tradeoff between the ASE and link reliability in multiuser MIMO HetNets.  To achieve the maximum ASE while guaranteeing a certain link reliability, we develop efficient algorithms to find the optimal BS densities. It is shown that as the link reliability requirement increases, more BSs and more tiers should be deactivated.
\end{abstract}

\vspace{-0.5cm}
\begin{IEEEkeywords}
\vspace{-0.5cm} Poisson point process, MIMO heterogeneous networks, area spectral efficiency, link reliability.
\end{IEEEkeywords}

\end{titlepage}

\baselineskip=24pt
\newpage

\section{Introduction}


Network densification via small cells -- resulting in heterogeneous cellular networks (HetNets) -- along with novel multiple antenna technologies are two of the key methods  for achieving a 1000x capacity increase in 5G networks \cite{Andrews14}.
By densely deploying different types of small cells overlaid on the existing macrocell network, spatial reuse and thus area spectral efficiency (ASE) can be significantly improved and uniform coverage can be provided \cite{Hwang13,Ghosh12,Bartoli14}.  Independently, multiple antenna technology, known as MIMO, is playing an increasingly important role in 4G cellular networks, and is expected to be even more essential for meeting 5G target data rates.  One key such technique is multiuser MIMO, which allows a base station (BS) with many antennas to communicate simultaneously with numerous mobile units each with a very small number of antennas.  Although multiuser MIMO -- also known as space division multiple access (SDMA) -- has been known for quite some time and previous implementation efforts have been relatively disappointing, enthusiasm has been recently renewed, as seen in the extensive recent literature on ``massive MIMO'' \cite{Rusek13}, as well as the very recent 3GPP standardization of full-dimension (FD) MIMO, which can support 64 antennas in a 2D array at the BS to communicate simultaneously with 32 mobile terminals \cite{Kim14}.

\subsection{Related Work}

A key result from early HetNet analysis was the derivation of the signal-to-interference-plus-noise (SINR) distribution, also known as the coverage or outage probability, where the HetNet is characterized by randomly placed base stations forming $K$ tiers, each tier distinguished by a unique transmit power and \emph{density}, i.e. the average number of BSs per unit area  \cite{Jo12,Dhillon12}.  A key resulting observation was that the signal-to-interference ratio (SIR) distribution is invariant to the BS densities, as long as the mobile connects to the BS providing the strongest received signal power. This \emph{SIR invariance} property means that cell densification does not degrade the link reliability, and so the area spectral efficiency (ASE) of the network can be increased indefinitely by deploying more BSs.  These early papers have resulted in a flurry of follow on work, e.g.
\cite{Singh13,Cheung12,Renzo13,Dhillon13_LA,DiRenzo14,DiRenzo15}, and see \cite{ElSawy13} for a survey.

Multi-antenna transmissions bring significant additional complexity to HetNet analysis, primarily due to the complexity of the  random matrix channel.  As shown in \cite{Dhillon13}, the invariance property may be lost in multi-antenna HetNets, i.e., the outage probability will increase as the BS density increases. This is mainly because the distributions of both the signal and interference depend on the number of BS antennas and the adopted multi-antenna transmission strategy of each BS. The work \cite{Dhillon13} relied on stochastic ordering to compare different multi-antenna techniques, but such a method cannot be used for quantitative analysis, since the SINR and SIR distributions were not provided. That work was extended to incorporate load balancing and thus the achievable rate in \cite{Gupta14}.  Other notable efforts on MIMO HetNets include work limited to two tiers  \cite{Chandrasekhar09,Adhikary15}, and the analysis in \cite{Heath13}, which focused on the interference distribution.
In this paper, we attempt to solve a more fundamental problem, namely the determination of the SIR distribution in a MIMO HetNet, particularly a multiuser MIMO HetNet, and the resulting effect of the BS density on the network performance.  By ``performance'', we are interested in both link reliability -- i.e. the success probability -- and the sum throughput, which we characterize per unit area as the ASE.

The link reliability vs. ASE tradeoff discussed in this paper is related to the notion of ``transmission capacity'' in wireless ad hoc networks \cite{Weber05,Weber12}, with related multi-antenna results such as \cite{Hunter08,Govindasamy07,Jindal11,Kountouris12}. In wireless ad hoc networks, to maximize the spatial throughput, i.e. ASE, while guaranteeing the link reliability at a certain SINR, the density of transmitters cannot exceed a certain value, which is called the \emph{transmission capacity} \cite{Weber12}. Naturally, allowing more simultaneous transmitters will increase the spatial reuse efficiency, but the interference to the receivers will become higher and so the SINR and thus link reliability will decrease, similar to the ASE vs. link reliability tradeoff studied in this paper.

\subsection{Contributions}

The main contributions of this paper are summarized as follows.

\subsubsection{Tractable Expressions of the Success Probability for Downlink Multiuser MIMO HetNets}

An exact expression of the success probability of the typical user in a general $K$-tier multiuser MIMO HetNet with zero-forcing (ZF) precoding and perfect channel state information (CSI) is derived. This expression has a simple structure and can be utilized for performance evaluation and system design of general MIMO HetNets. Moreover, based on this exact expression, an asymptotic (large SIR) expression of the success probability is also provided, which has a simple and symmetric form with respect to the network parameters.

\subsubsection{Key Properties of the Success Probability}

Based on these tractable expressions, we explicitly show the effect of the BS densities on the success probability, i.e., the link reliability. First, we prove that increasing the BS density of one tier will either increase or decrease the link reliability. Thus, the SIR invariance property no longer holds in MIMO HetNets. Second, we prove that the maximum link reliability of the network is achieved by only activating one tier of BSs. This is a surprising result, as nowadays a common intuition is that deploying more and more BSs will improve the network performance. Moreover, we show that such phenomena result from the fact that different tiers in MIMO HetNets have different influences on the link reliability, and the overall link reliability is an average of users in different tiers.

\subsubsection{ASE vs. Link Reliability Tradeoff}

Since adjusting the BS densities will affect both the link reliability and ASE, a tradeoff is needed to balance these two aspects. Thus, we propose efficient algorithms to find the optimal BS densities that can achieve the maximum ASE while guaranteeing a certain link reliability requirement. If the required link reliability is low, all the BSs should be activated to achieve the maximum ASE. On the contrary, if the required link reliability becomes higher, more BSs should be deactivated. The extreme case is that to achieve the highest link reliability, only a single tier of BSs should be active. Therefore, care is needed when densifying a HetNet with multi-antenna BSs.

\subsection{Paper Organization and Notation}

The rest of the paper is organized as follows. Section~\ref{Sec:SystemModel} presents the system model. Section~\ref{Sec:PsAnalysis} derives the expressions of the success probability. In Section~\ref{Sec:Tradeoff}, we investigate the tradeoff between the ASE and link reliability. Finally, Section~\ref{Sec:Conclusions} concludes the paper.

In this paper, vectors and matrices are denoted by hold-face lower-case and upper-case letters, respectively. The $M\times M$ identity matrix is denoted by $\mathbf{I}_M$. We use $\left(\cdot\right)^{T}$ and $\left(\cdot\right)^{*}$ to denote transpose and conjugate transpose, respectively.  The Euclidean norm of the vector $\mathbf{x}$ is $\left\Vert \mathbf{x}\right\Vert$, while $\left\Vert \mathbf{A}\right\Vert _{1}$ is the $L_1$ induced matrix norm, i.e., $\left\Vert \mathbf{A}\right\Vert _{1}=\max_{1\leq j\leq n}\sum_{i=1}^{m}\left|a_{ij}\right|$ for $\mathbf{A}\in\mathbb{R}^{m\times n}$. The expectation is denoted as $\mathbb{E}\left[\cdot\right]$. The notation $X \stackrel{d}{\sim} Y$ means that $X$ is distributed as $Y$. The normalized sinc function is defined by ${\rm sinc}\left(\delta\right)=\frac{\sin\left(\pi \delta\right)}{\pi \delta}$. Finally, $\mathcal{O}\left(\cdot\right)$ is the Landau's notation, where $f\left(x\right)=\mathcal{O}\left(g\left(x\right)\right)$ as $x\rightarrow x_0$ means that there are constants $C>0$ and $\varepsilon>0$, such that for $0<\left|x-x_{0}\right|<\varepsilon$, $\left|f\left(x\right)\right|\leq C\left|g\left(x\right)\right|$. The asymptotic equivalence is denoted as $\sim$, where $f\left(x\right)\sim g\left(x\right)$ as $x\rightarrow\infty$ means that $f\left(x\right)/g\left(x\right)$ tends to unity as $x\rightarrow\infty$. 


\section{System Model} \label{Sec:SystemModel}

In this section, we will first introduce a stochastic model for downlink multiuser MIMO HetNets, and then the main performance metrics considered in the paper will be presented.

\subsection{A Tractable Model for Downlink MIMO HetNets}

\begin{figure}
    \begin{center}
    \scalebox{0.7}{\includegraphics{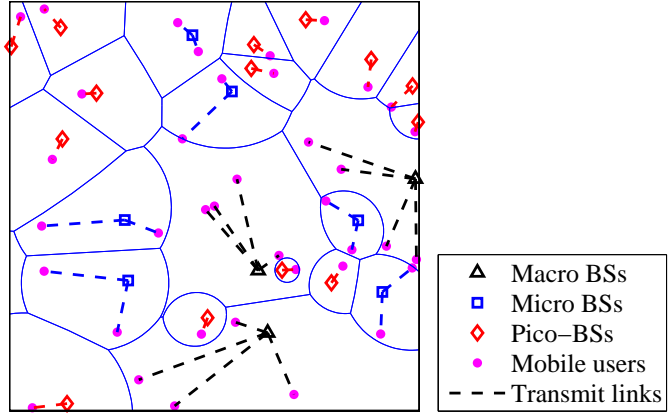}}
    \end{center}
    \caption{A demonstration of a 3-tier HetNet, in which each macro BS has 8 antennas and serves 4 users, each micro BS has 4 antennas and serves 2 users, and each pico BS has a single antenna and serves 1 user.}
    \label{fig:System_Model}
\end{figure}

We consider a downlink cellular network consisting of $K$ different tiers of BSs, indexed by $\mathcal{K}=\left\{1,2,\ldots,K\right\}$. In the $k$th tier, the BSs are spatially distributed according to a homogeneous Poisson point process (PPP) \cite{Stoyan96} in $\mathbb{R}^2$ of density $\lambda_k$, denoted by $\Psi_k$. Each BS in the $k$th tier has transmit power $P_k$, with $M_k$ antennas, and serves $U_k$ ($U_k\leq M_k$) users at each time slot, i.e., intra-cell SDMA is considered. Note that when $U_k=1$, $\forall k\in\mathcal{K}$, the network becomes a TDMA HetNet. Meanwhile, we assume that mobile users are distributed as a homogeneous PPP $\Phi_u$ of density $\lambda_u$, which is independent of $\Psi_k$, $\forall k\in\mathcal{K}$. Each user has a single receive antenna, and is associated with one BS\footnote{Note that our analysis also applies to single user MIMO, where each user in the $k$th tier has $U_k$ uncorrelated receive antennas, and BSs in the $k$th tier apply equal power allocation to the $U_k$ streams. Then the success probability to the typical user analyzed in this paper becomes the success probability per stream.}. We assume the network is fully-loaded ($\lambda_u\gg\lambda_k,\forall k\in\mathcal{K}$), i.e., there are at least $U_k$ users in each cell in the $k$th tier, and each BS will randomly choose $U_k$ users to serve at each time slot. Note that this is a common assumption in the analysis of random cellular networks \cite{Singh13,Gupta14}, and the analysis of the non-fully-loaded network is left to our future work. An example of a 3-tier multiuser MIMO HetNet is shown in Fig.~\ref{fig:System_Model}.

The network is open access, which means a user is allowed to access any BS in any tier. Particularly, each user will listen to the pilot signals from different BSs, and measure the long-term average received power. We will consider a commonly adopted cell association rule \cite{Jo12}, i.e., the user will be associated with the $k$th tier if\footnote{In the user association procedure, the first antenna normally uses the total transmission power of a BS to transmit reference signals for biased received power determination according to the LTE standard \cite{Ghosh10}.}
\begin{equation}
    k=\arg\max_{j\in\mathcal{K}} P_j B_j r_j^{-\alpha},
\end{equation}
where $r_j$ denotes the minimal distance from a user to its nearest BS in the $j$th tier, $\alpha>2$ is the path loss exponent\footnote{Note that in this paper, we assume the same path loss exponent in the network. However, the analysis can be extended to the model in which different tiers have different path loss exponents (e.g., \cite{Jo12}).}, and $B_j$ is the bias factor of the $j$th tier, which is used for load balancing \cite{Jo12,ElSawy13}. It is shown in \cite{Jo12} that the typical user will be associated with the $k$th tier with probability
\begin{equation}
    A_k = \frac{\lambda_k P_k^\delta B_k^\delta}{A},
\end{equation}
where $\delta\triangleq \frac{2}{\alpha}$ and $A=\sum_{j=1}^K \lambda_j P_j^\delta B_j^\delta$.

For each BS, equal power allocation is assumed in this paper, i.e., the BS in the $k$th tier will equally allocate $P_k$ to its $U_k$ users. Furthermore, we adopt ZF precoding and assume perfect CSI at each BS\footnote{ZF is commonly used as the transmission strategy for multiuser MIMO  in both the literature and industry, due to its simplicity and fairly well performance. It will be interesting to investigate how the findings in this paper will change if other precoding methods are used. However, this is left to our future work.}. Therefore, the received power of the typical user at origin $o$ from a BS located at $x\in\Psi_k$ is given by
\begin{equation}
    P\left(x\right) = \frac{P_k}{U_k} g_{x,k} \left\Vert x\right\Vert ^{-\alpha},
\end{equation}
where $g_{x,k}$ is the channel gain, and its distribution depends on whether the BS is the home BS (i.e., the associated BS) or the interfering BS. Assuming Rayleigh fading channels, it is shown in \cite{Heath11} that $g_{x,k}$ is gamma distributed with the shape parameter $M_k-U_k+1$ if the BS at $x$ is the home BS, i.e., $g_{x,k}\stackrel{d}{\sim}{\rm Gamma}\left(D_k,1\right)$, where $D_k\triangleq M_k-U_k+1$, while it can be approximated as $g_{x,k}\stackrel{d}{\sim}{\rm Gamma}\left(U_k,1\right)$ if the BS is the interfering BS in the $k$th tier. This approximation has been commonly adopted in performance analysis of multiuser MIMO networks, e.g., \cite{Dhillon13,Hosseini14,Chandrasekhar09,Kountouris12}. In addition, the gamma distribution is a good approximation for the channel gain distributions for a large class of multi-antenna transmission schemes \cite{Heath13}, and thus the analysis in this paper can be extended to more general settings.


In this paper, we assume that  universal frequency reuse is applied, which implies that the user will not only receive the information signal from its home BS, but also suffer interference from all the other BSs. The resulting SIR of the typical user, served by the BS at $x_0$ in the $k$th tier, is given by
\begin{equation} \label{eq:SINR_def}
    {\rm SIR}_{k}= \frac{\frac{P_{k}}{U_{k}}g_{x_{0},k}r_{k}^{-\alpha}} {\sum_{j=1}^{K}\sum_{x\in\Psi_{j}\backslash\left\{ x_{0}\right\}} \frac{P_{j}}{U_{j}}g_{x,j}\left\Vert x\right\Vert ^{-\alpha}}.
\end{equation}
Note that we consider SIR, rather than SINR, in this paper, as the SIR distribution captures the key tradeoffs in reasonably dense deployments.

\begin{remark}
    In summary, the most important parameters of the network are the per-tier parameters: $\left\{\lambda_k,P_k,B_k,M_k,U_k\right\}$. Compared with existing works, the $K$-tier HetNet model in this paper includes intra-cell multi-antenna transmissions, which generalizes the SISO HetNet model (where $M_k=U_k=1$ for all $k\in\mathcal{K}$) \cite{Jo12}. Based on this model, the effects of the multiple antennas and the number of served users can be evaluated. Meanwhile, the derivation and analysis will be more difficult, as both the signal and interfering channel gains become gamma distributed, while in SISO HetNets, all the channel gains are exponentially distributed.
\end{remark}

\subsection{Performance Metrics}

In this paper, we consider two performance metrics. The first one is the success probability, also called the coverage probability \cite{Andrews11,Jo12}, which measures the reliability of the typical transmission link. The definition of the success probability for the typical user is given by
\begin{equation} \label{eq:Ps_def}
    p_{\rm s}= \mathbb{P}\left({\rm SIR}\geq\hat\gamma\right),
\end{equation}
where $\hat\gamma$ is the SIR threshold. Since the success probability is different if the typical user is associated with different tiers, the overall success probability in \eqref{eq:Ps_def} can be written as in \cite{Jo12} as
\begin{equation} \label{eq:Psk_def}
    p_{\rm s}= \sum_{k=1}^K A_k p_{\rm s} \left(k\right),
\end{equation}
where $p_{\rm s} \left(k\right)$ is the success probability when the user is served by the BS in the $k$th tier.

On the other hand, to describe the network capacity, we use the ASE as the metric\footnote{Note that in some literature, e.g., \cite{Andrews11} and \cite{Wang14}, the Shannon throughput of a cell, defined by $\mathbb{E}\left[\log_2\left(1+{\rm SIR}\right)\right]$, is used as the performance metric. The extension of our results to this case will be left to future work.}, defined by \cite{Singh13,Baccelli06,Cheung12} as
\begin{equation} \label{eq:ASE_def}
    {\rm ASE}=\sum_{k=1}^K \lambda_k U_k p_{\rm s} \left(k\right) \log_2\left(1+\hat\gamma\right),
\end{equation}
for which the unit is bit/s/Hz/${\rm m}^2$. Note that ASE measures the total data rate in a unit area normalized by the bandwidth, and it can directly tell the capacity gain by cell densification.
The ASE defined in \eqref{eq:ASE_def} applies to the fully-loaded networks. Note that if the network is not fully-loaded, the inactive BSs will increase the SIR and thus the success probability, but decrease the achieved ASE since there are fewer active links per unit area \cite{Dhillon13_LA}.

The success probability and ASE are two different aspects of a communication system, and both of them are fundamental metrics. A higher ASE means the network can support more users, i.e., with a higher spatial reuse efficiency, while a higher success probability implies that the transmission links are more reliable, i.e., a better quality of experience (QoE). While previous studies revealed the invariance property of the success probability in SISO HetNets, in this paper, we will investigate the interplay of these two metrics in more general multiuser MIMO HetNets. Since the ASE depends on $p_{\rm s} \left(k\right)$, we will first derive it in Section \ref{Sec:PsAnalysis}, and then investigate the tradeoff between ASE and $p_{\rm s}$ in Section \ref{Sec:Tradeoff}.

\section{Analysis of the Success Probability} \label{Sec:PsAnalysis}

In this section, we will first derive an exact expression of the success probability. Then, an asymptotic expression will be provided, which will expose its key properties in MIMO HetNets.

\subsection{The Exact Expression of $p_{\rm s}$}

To obtain the overall success probability $p_{\rm s}$ from \eqref{eq:Psk_def}, we need to first derive the per-tier values $p_{\rm s}\left(k\right)$. Assuming the typical user is served by the BS in the $k$th tier, then based on \eqref{eq:SINR_def}, $p_{\rm s}\left(k\right)$ is given by
\begin{equation} \label{eq:Psk_ori}
    p_{{\rm s}}\left(k\right)= \mathbb{P}\left(\frac{\frac{P_{k}}{U_{k}}g_{x_{0},k}r_{k}^{-\alpha}} {\sum_{j=1}^{K}\sum_{x\in\Psi_{j}\backslash\left\{ x_{0}\right\} }\frac{P_{j}}{U_{j}}g_{x,j}\left\Vert x\right\Vert ^{-\alpha}}\geq\hat{\gamma}\right).
\end{equation}
The main difficulty of the derivation is the gamma distributed channel gains $g_{x_{0},k}$ and $g_{x,j}$. There exist some previous works studying similar network models. While only a qualitative comparison between SDMA and TDMA was considered in \cite{Dhillon13}, the complicated expression of $p_{\rm s}$ derived in \cite{Gupta14} based on the Fa{\`a} di Bruno lemma obscured useful design insights. In this paper, we adopt a novel method and derive a more tractable expression for $p_{\rm s}$, as shown in Theorem \ref{Thm:Psk_General}. 

\begin{theorem} \label{Thm:Psk_General}
    The success probability of the typical user served by the BS in the $k$th tier is given by
    \begin{equation} \label{eq:Psk_Matrix}
        p_{s}\left(k\right)=A\left\Vert \mathbf{Q}_{D_{k}}^{-1}\right\Vert _{1},
    \end{equation}
    where $A=\sum_{j=1}^K \lambda_j P_j^\delta B_j^\delta$, $\delta=\frac{2}{\alpha}$, $D_{k}\triangleq M_{k}-U_{k}+1$, $\left\Vert \cdot\right\Vert _{1}$ is the $L_1$ induced matrix norm, and $\mathbf{Q}_{D_k}$ is the $D_k\times D_k$ lower triangular Toeplitz matrix, given by
    \begin{equation} \label{eq:Q}
    \mathbf{Q}_{D_{k}}=\left[\begin{array}{ccccc}
        q_{0}\\
        q_{1} & q_{0}\\
        q_{2} & q_{1} & q_{0}\\
        \vdots &  &  & \ddots\\
        q_{D_{k}-1} & \cdots & q_{2} & q_{1} & q_{0}
        \end{array}\right],
    \end{equation}
    in which
    \begin{equation} \label{eq:qi}
        q_{i} = \sum_{j=1}^{K}\lambda_{j}P_{j}^{\delta}B_{j}^{\delta} \frac{\Gamma\left(U_j+i\right)}{\Gamma\left(U_j\right)\Gamma\left(i+1\right)} \frac{\delta}{\delta-i} \left(\frac{U_{k}B_{k}}{U_{j}B_{j}}\hat{\gamma}\right)^{i}
        {}_{2}F_{1} \left(i-\delta,U_{j}+i;i+1-\delta;-\frac{U_{k}B_{k}}{U_{j}B_{j}}\hat{\gamma}\right),
    \end{equation}
    where ${}_{2}F_{1}\left(a,b;c;z\right)$ is the Gauss hypergeometric function \cite[p.~1005]{Table}.
\end{theorem}
\begin{IEEEproof}
    See Appendix \ref{App:Psk_General}.
\end{IEEEproof}

Note that expression \eqref{eq:Psk_Matrix} is similar to the expression of $p_{\rm s}$ of multi-antenna single-tier networks \cite{Li14}. However, compared with the result in \cite{Li14}, the expression in Theorem~\ref{Thm:Psk_General} is in a simpler form while dealing with more general cases. Furthermore, the derivation provided in Appendix~\ref{App:Psk_General} adopts a different method which is more tractable than the one in \cite{Li14}. Accordingly, the success probability of the typical user is given by
\begin{equation}
    p_{\rm s} = \sum_{k=1}^K \lambda_k P_k^\delta B_k^\delta \left\Vert \mathbf{Q}_{D_{k}}^{-1}\right\Vert _{1}.
\end{equation}

The success probability of the single-tier network is given as a special case in the following corollary.
\begin{corollary} \label{Col:SingleTier}
    The success probability of the typical user in the single-tier network where each BS has $M$ antennas and serves $U$ users is given by
    \begin{equation} \label{eq:Ps_singletier}
        p_{s}\left(k\right)=\left\Vert \mathbf{\tilde{Q}}_{D}^{-1}\right\Vert _{1},
    \end{equation}
    where  $D = M-U+1$, and $\mathbf{\tilde{Q}}_{D}$ has the same structure as $\mathbf{Q}_{D_k}$ in \eqref{eq:Q}, with the elements $\tilde{q}_i$ given by
    \begin{equation} \label{eq:qi_singletier}
        \tilde{q}_i=\frac{\Gamma\left(U+i\right)}{\Gamma\left(U\right)\Gamma\left(i+1\right)} \frac{\delta}{\delta-i}\hat{\gamma}^{i} {}_{2}F_{1}\left(i-\delta,U+i;i+1-\delta;-\hat{\gamma}\right).
    \end{equation}
\end{corollary}
It shows in Corollary~\ref{Col:SingleTier} that the success probability of the typical user is independent to the BS density, which indicates that the SIR-invariance property still holds in SDMA homogeneous cellular networks, which is consistent with SISO networks \cite{Andrews11}.

Theorem~\ref{Thm:Psk_General} provides a tractable expression that can be easily evaluated numerically. However, it is still in a complicated form, which makes it difficult to directly observe the effects of different system parameters. To overcome this difficulty, in the next subsection, we will provide an asymptotic result of $p_{\rm s}$ as the SIR threshold becomes large.

\subsection{The Asymptotic Expression of $p_{\rm s}$ and Its Properties}

In this subsection, we will first derive an asymptotic expression of $p_{\rm s}$ as the SIR threshold is large, i.e., $\hat{\gamma}\rightarrow \infty$. Then, we will provide some basic properties of $p_{\rm s}$, which can help to better understand multiuser MIMO HetNets.

The Taylor expansion of the coefficients $q_i$ in \eqref{eq:qi} gives
\begin{equation} \label{eq:qi_Asy}
    q_{i}= \sum_{j=1}^{K}\lambda_{j}P_{j}^{\delta}B_{j}^{\delta} \left(\frac{U_{k}B_{k}}{U_{j}B_{j}}\hat{\gamma}\right)^{\delta} \frac{\delta}{\delta-i}\frac{\Gamma\left(i+1-\delta\right)\Gamma\left(U_{j} +\delta\right)} {\Gamma\left(i+1\right)\Gamma\left(U_{j}\right)} +\mathcal{O}\left(\frac{1}{\hat{\gamma}^{U_{j}}}\right).
\end{equation}
Then, under the assumption that $\hat\gamma\rightarrow\infty$, the asymptotic expression of the success probability is given in the following theorem.

\begin{theorem} \label{Thm:Psk_Asy}
    The asymptotic expression of $p_{\rm s}\left(k\right)$ as $\hat\gamma\rightarrow\infty$ is given by
    \begin{equation} \label{eq:Psk_Asy}
        p_{{\rm s}}\left(k\right)\sim A\hat{\gamma}^{-\delta} {\rm sinc}\left(\delta\right) \frac{\left(U_{k}B_{k}\right)^{-\delta} \frac{\Gamma\left(D_{k}+\delta\right)}{\Gamma\left(D_{k}\right)}} {\sum_{j=1}^{K}\lambda_{j}\left(\frac{P_{j}}{U_{j}}\right)^{\delta} \frac{\Gamma\left(U_{j}+\delta\right)}{\Gamma\left(U_{j}\right)}}.
    \end{equation}
\end{theorem}
\begin{IEEEproof}
    See Appendix \ref{App:Psk_Asy}.
\end{IEEEproof}

Based on Theorem~\ref{Thm:Psk_Asy}, the success probability of the typical user is given by
\begin{equation} \label{eq:Ps_Asy}
    p_{{\rm s}}=\sum_{k=1}^{K}A_{k}p_{{\rm s}}\left(k\right)\sim \hat{\gamma}^{-\delta}{\rm sinc}\left(\delta\right) \frac{\sum_{k=1}^{K}\lambda_{k}\left(\frac{P_{k}}{U_{k}}\right)^{\delta} \frac{\Gamma\left(D_{k}+\delta\right)}{\Gamma\left(D_{k}\right)}} {\sum_{j=1}^{K}\lambda_{j}\left(\frac{P_{j}}{U_{j}}\right)^{\delta} \frac{\Gamma\left(U_{j}+\delta\right)}{\Gamma\left(U_{j}\right)}}.
\end{equation}

\begin{figure}
    \begin{center}
    \scalebox{0.7}{\includegraphics{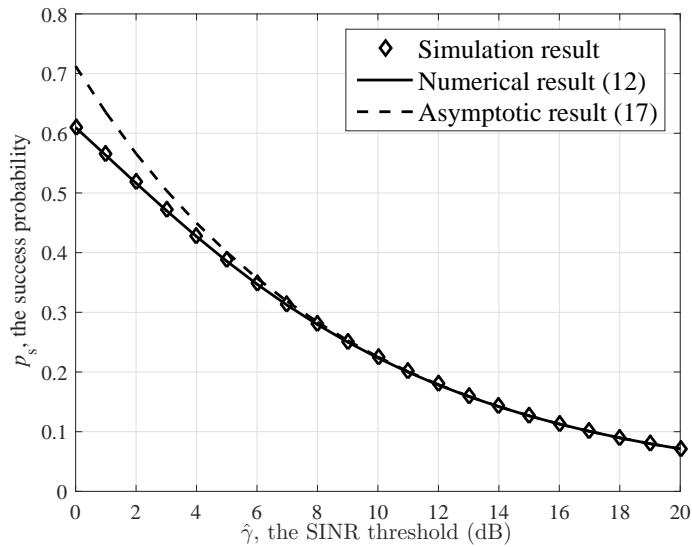}}
    \end{center}
    \caption{The success probability with different SIR threshold, with $\alpha=4$, $\left[\lambda_1,\lambda_2,\lambda_3\right]=\left[1, 5, 10\right]\times10^{2}$ per ${\rm km}^2$, $\left[M_1,M_2, M_3\right]=\left[8, 4, 1\right]$, $\left[U_1,U_2, U_3\right]=\left[4, 2, 1\right]$, $\left[P_1,P_2, P_3\right]=\left[6.3, 0.13, 0.05\right]$W and $B_k=1/U_k$.}
    \label{fig:Ps_Asy}
\end{figure}

In Fig. \ref{fig:Ps_Asy}, we compare the asymptotic result in \eqref{eq:Ps_Asy} with the exact numerical result in Theorem~\ref{Thm:Psk_General}, as well as the simulation result. We consider a 3-tier HetNet, and the parameters are provided in the caption of the figure. From Fig. \ref{fig:Ps_Asy}, we can find that the numerical result in Theorem~\ref{Thm:Psk_General} is exactly the same as the simulation result. Moreover, the asymptotic result in \eqref{eq:Ps_Asy} provides as an upper bound when the SIR threshold $\hat\gamma$ is not large, but the gap becomes smaller when $\hat\gamma$ increases. This result confirms the effectiveness of the asymptotic result when the SIR threshold is large. For example, in this setting, the asymptotic result is almost the same as the exact result when $\hat\gamma \gtrsim 5$ dB, and above 10 dB, is indistinguishable. Note that MIMO techniques will enhance the data rate at relatively high SIRs, so $\hat\gamma \gtrsim 5$ dB is practical in MIMO HetNets.

From \eqref{eq:Ps_Asy}, we can see a clear relation between the success probability and each parameter. It is a perfectly symmetric structure, since both the numerator and denominator are summations of each tier's density $\left\{\lambda_k\right\}$\footnote{In this paper, we use $\left\{\lambda_k\right\}$ to denote the set $\left\{\lambda_k:k\in\mathcal{K}\right\}$, representing the BS densities of all the tiers, while $\lambda_k$ is used to denote the BS density of the $k$th tier.}, the transmit power to each user $\left\{\frac{P_k}{U_k}\right\}$, and the effects of link array gain $\left\{D_k\right\}$ and multiplexing gain $\left\{U_k\right\}$, respectively. Interestingly, this expression bears a similar form with the one in \cite{Dhillon12}, which considered SISO HetNets. Based on \eqref{eq:Ps_Asy}, we can easily investigate the effects of different parameters on the success probability in the general MIMO HetNets. Due to space limitations, in this paper, we will focus on how the BS densities $\left\{\lambda_k\right\}$ affect the success probability, and present two important properties. For convenience, we rewrite \eqref{eq:Ps_Asy} in the vector form by defining the column vectors $\boldsymbol{\lambda}=\left[\lambda_{1},\lambda_{2},\ldots,\lambda_{K}\right]^{T}$, $\mathbf{c}=\left[c_1,c_2,\ldots,c_K\right]^{T}$ where $c_i=\left(\frac{P_{i}}{U_{i}}\right)^{\delta} \frac{\Gamma\left(D_{i}+\delta\right)}{\Gamma\left(D_{i}\right)}$, and $\mathbf{d}=\left[d_1,d_2,\ldots,d_K\right]^{T}$ where $d_i=\left(\frac{P_{i}}{U_{i}}\right)^{\delta} \frac{\Gamma\left(U_{i}+\delta\right)}{\Gamma\left(U_{i}\right)}$. Then, \eqref{eq:Ps_Asy} is equivalent to
\begin{equation} \label{eq:Ps_Asy_Vector}
    p_{{\rm s}}\sim\hat{\gamma}^{-\delta}{\rm sinc}\left(\delta\right) \frac{\mathbf{c}^{T}\boldsymbol{\lambda}}{\mathbf{d}^{T}\boldsymbol{\lambda}}.
\end{equation}

\subsubsection{Monotonicity of $p_{\rm s}$ with respect to $\lambda_k$}

Based on \eqref{eq:Ps_Asy_Vector}, we can obtain the following result.
\begin{lemma} \label{Lemma:Ps_Monotonic}
    When the SIR threshold $\hat{\gamma}$ is large, if the ratio $\frac{c_{k}}{d_{k}}= \frac{\Gamma\left(D_{k}+\delta\right)/\Gamma\left(D_{k}\right)} {\Gamma\left(U_{k}+\delta\right)/\Gamma\left(U_{k}\right)}$ is the same for all the tiers, $p_{\rm s}$ is invariant with $\lambda_k$, $\forall k\in \mathcal{K}$; Otherwise, $p_{\rm s}$ is monotonic with respect to $\lambda_k$, i.e., $p_{\rm s}$ will either increase or decrease as $\lambda_k$ increases.
\end{lemma}
\begin{IEEEproof}
    See Appendix~\ref{App:Ps_Monotonic}.
\end{IEEEproof}

This result explicitly shows that there is, in general, no invariance property for $p_{\rm s}$ in MIMO HetNets. Deploying more BSs of one tier will either increase or decrease the success probability. The invariance property that $p_{\rm s}$ is independent of the BS densities in SISO HetNets is only a special case, where $\frac{c_{i}}{d_{i}}=1$ for all the tiers. Therefore, we should carefully consider how the success probability will be affected when densifying the HetNet with multi-antenna BSs.

\subsubsection{The maximum $p_{\rm s}$ of the network}

In this part, we will determine the maximum $p_{\rm s}$ that can be achieved in a given MIMO HetNet. This is equivalent to considering the following optimization problem:
\begin{eqnarray} \label{eq:OptiPs_Asy}
    \underset{\left\{ \lambda_{k}\right\} }{\text{maximize}} &  p_{{\rm s}}  \\
    \text{subject to} &  0\leq\lambda_k \leq \lambda_k^{\max}, \forall k\in\mathcal{K}, \nonumber
\end{eqnarray}
where $\lambda_k$ can be regarded as the active BS density of the $k$th tier, and $\lambda_k^{\max}$ is the actual BS density. The solution of this optimization problem is provided in Lemma~\ref{Lemma:MaxPs_Asy}.

\begin{lemma} \label{Lemma:MaxPs_Asy}
    The maximum $p_{\rm s}$ with respect to $\left\{\lambda_k\right\}$ in MIMO HetNets is given by
    \begin{equation} \label{eq:Ps_Asy_Max}
        p_{{\rm s}}^{\max}=p_{\rm s}\left(k\right)\quad\text{for }k=\arg\max_{j}\frac{\Gamma\left(D_{j}+\delta\right)/\Gamma\left(D_{j}\right)} {\Gamma\left(U_{j}+\delta\right)/\Gamma\left(U_{j}\right)},
    \end{equation}
    and the optimal BS density is
    \begin{equation}
        \lambda_{k}^{\star}\begin{cases}
        \in\left(0,\lambda_{k}^{\max}\right] & k=\arg\max_{j} \frac{\Gamma\left(D_{j}+\delta\right)/\Gamma\left(D_{j}\right)} {\Gamma\left(U_{j}+\delta\right)/\Gamma\left(U_{j}\right)},\nonumber\\
        =0 & k\neq\arg\max_{j} \frac{\Gamma\left(D_{j}+\delta\right)/\Gamma\left(D_{j}\right)} {\Gamma\left(U_{j}+\delta\right)/\Gamma\left(U_{j}\right)}. \nonumber
        \end{cases}
    \end{equation}
\end{lemma}
\begin{IEEEproof}
    See Appendix~\ref{App:MaxPs}.
\end{IEEEproof}

From Lemma~\ref{Lemma:MaxPs_Asy}, we see that when the numbers of antennas and the numbers of served users are determined, i.e., when $\left\{D_k,U_k\right\}$ are determined, then no matter how to change the BS densities, the success probability of the network cannot exceed the value in \eqref{eq:Ps_Asy_Max}. Furthermore, the maximum $p_{\rm s}$ is achieved by only activating one tier of BSs, which has the largest value of $\frac{\Gamma\left(D_{i}+\delta\right)/\Gamma\left(D_{i}\right)} {\Gamma\left(U_{i}+\delta\right)/\Gamma\left(U_{i}\right)}$.

The two properties above have shown the effect of the BS densities on the success probability in a general MIMO HetNet. In the next section, we will provide a more comprehensive investigation and jointly analyze the effects of the BS densities on the success probability and ASE.

\section{ASE and Link Reliability Tradeoff} \label{Sec:Tradeoff}

In this section, we will investigate the effects of the BS densities on both the link reliability and ASE, and show there is a tradeoff between these two metrics. Moreover, efficient algorithms will be proposed to find the optimal BS densities.   To start with, we will first consider a special case where all the BSs serve the same number of users. The purpose of investigating such a special case is to reveal the physical insight of the tradeoff. Then, we will show the tradeoff in general MIMO HetNets using the asymptotic results provided in the last section.

\subsection{ASE and Link Reliability Tradeoff in $U$-SDMA Networks}

In this subsection, we will consider the tradeoff between the ASE and link reliability in the $U$-SDMA network \cite{Li13}, where each BS serves $U$ ($U_k=U\leq M_k$ for all $k\in\mathcal{K}$) users. By limiting each BS to serve the same number of users, exact expressions of $p_{\rm s}$ and ASE are available. Such a network still possesses the key characteristics of the MIMO HetNet. For example, when $U=1$, the network becomes a TDMA HetNet. For simplicity, we assume $B_k=1$ for $\forall k\in\mathcal{K}$. Then $p_{\rm s}\left(k\right)$ can be directly obtained from Theorem \ref{Thm:Psk_General}, as given in the following corollary.

\begin{corollary} \label{col:USDMA}
    The success probability of the typical user served by the $k$th tier in the unbiased $U$-SDMA HetNet is given by
    \begin{equation}
        p_{{\rm s}}\left(k\right)  =\left\Vert \tilde{\mathbf{Q}}_{D_{k}}^{-1}\right\Vert _{1},
    \end{equation}
    where $\tilde{\mathbf{Q}}_{D_{k}}$ has the same structure as ${\mathbf{Q}}_{D_{k}}$ in \eqref{eq:Q}, with the elements $\tilde{q}_i$ given in \eqref{eq:qi_singletier}.
\end{corollary}

Therefore, the success probability of the typical user is given by
\begin{equation} \label{eq:Ps_USDMA}
    p_{{\rm s}}= \sum_{k=1}^{K}A_{k}p_{{\rm s}}\left(k\right) = \frac{\sum_{k=1}^{K}\left(\frac{P_{k}}{U}\right)^{\delta}p_{{\rm s}}\left(k\right)\lambda_{k}}{\sum_{k=1}^{K}\left(\frac{P_{k}}{U}\right)^{\delta}\lambda_{k}},
\end{equation}
and its vector form is given by
\begin{equation}\label{eq:Ps_USDMA_Vector}
    p_{{\rm s}}=  \frac{\tilde{\mathbf{c}}^{T}\boldsymbol{\lambda}}{\tilde{\mathbf{d}}^{T}\boldsymbol{\lambda}},
\end{equation}
where the $k$th elements of $\tilde{\mathbf{c}}$  and $\tilde{\mathbf{d}}$ are $\tilde{c}_k=\left(\frac{P_{k}}{U}\right)^{\delta}p_{{\rm s}}\left(k\right)$ and $\tilde{d}_k=\left(\frac{P_{k}}{U}\right)^{\delta}$, respectively.

From Corollary~\ref{col:USDMA}, we have the following observations.
\begin{lemma} \label{Lemma:Psk_USDMA}
    $p_{\rm s}\left(k\right)$ in unbiased $U$-SDMA HetNets has the following properties:
    \begin{itemize}
      \item $p_{{\rm s}}\left(k\right)$ is independent of the BS densities $\left\{\lambda_k\right\}$ and the transmit power $\left\{P_k\right\}$.
      \item $p_{\rm s}\left(i\right)\geq p_{\rm s}\left(j\right)$ if and only if $M_i\geq M_j$, and the equality holds only if $M_i = M_j$.
    \end{itemize}
\end{lemma}
This result means that the tier with more antennas at each BS provides higher link reliability, since the interference suffered by each user has the same distribution in unbiased $U$-SDMA HetNets, while more transmit antennas will provide a higher diversity and array gain. While $p_{{\rm s}}\left(k\right)$ is independent of both $\left\{\lambda_k\right\}$ and $\left\{P_k\right\}$, the overall success probability $p_{\rm s}$ in \eqref{eq:Ps_USDMA} depends on $\left\{\lambda_k,P_k\right\}$, and $p_{\rm s}$ can be regarded as a weighted sum of $p_{{\rm s}}\left(k\right)$ for $k\in\mathcal{K}$. From \eqref{eq:Ps_USDMA_Vector}, the network success probability has the following properties:
\begin{lemma} \label{Lemma:Ps_USDMA}
    $p_{\rm s}$ in unbiased $U$-SDMA HetNets has the following properties:
    \begin{itemize}
      \item $p_{{\rm s}}$ is monotonic with respect to $\lambda_k$ for $\forall k\in\mathcal{K}$.
      \item The maximum $p_{\rm s}$ is achieved by only activating one tier of BSs which has the largest number of antennas, i.e.,
          \begin{equation} \label{eq:Ps_Max_USDMA}
            p_{{\rm s}}^{\max}=p_{\rm s}\left(k\right)\quad\text{for }k=\arg\max_{j}M_{j}.
          \end{equation}
    \end{itemize}
\end{lemma}
\begin{IEEEproof}
    Since \eqref{eq:Ps_USDMA_Vector} has the same structure with \eqref{eq:Ps_Asy_Vector}, Lemmas \ref{Lemma:Ps_Monotonic} and \ref{Lemma:MaxPs_Asy} can be applied. Moreover, we have $\frac{\tilde{c}_k}{\tilde{d}_k}=p_{\rm s}\left(k\right)$ in this case, and from Lemma~\ref{Lemma:Psk_USDMA}, we know the maximum $p_{\rm s}\left(k\right)$ is determined by $M_k$.
\end{IEEEproof}

By now, it is clear that the success probabilities in different tiers ($p_{\rm s}\left(k\right)$) are different, and $p_{\rm s}$ is an average of each tier's performance. Thus, the densification of different tiers will have different effects on $p_{\rm s}$. In particular, increasing the BS density of the tier with a lower $p_{\rm s}\left(k\right)$ will pull down the overall $p_{\rm s}$.

On the other hand, increasing the BS density will increase the ASE, which can be found in its expression:
\begin{equation} \label{eq:ASE_USDMA}
    {\rm ASE}=U\log_2\left(1+\hat\gamma\right)\sum_{k=1}^K \lambda_k  p_{\rm s} \left(k\right).
\end{equation}
Note that since $p_{\rm s}\left(k\right)$ is independent of the BS densities, \eqref{eq:ASE_USDMA} shows that activating as many BSs as possible can achieve a higher ASE. Combining the above results, we have the following two conflicting aspects:
\begin{itemize}
  \item To achieve the maximum ASE, it is optimal to activate all the BSs, but in this case $p_{{\rm s}}$ may be low.
  \item To achieve the maximum $p_{{\rm s}}$, activating only the tier with the largest number of antennas is optimal, but the ASE will be low.
\end{itemize}
Thus we need to investigate the tradeoff between ASE and link reliability. We are interested in maximizing the achievable ASE given a requirement on $p_{{\rm s}}$, formulated as the following problem.
\begin{eqnarray} \label{eq:Problem_ori}
  \mathcal{P}_o : \underset{\left\{\lambda_k\right\}}{\text{maximize}} & {\rm ASE} \\
  \text{subject to} & p_{\rm s} \geq \Theta, \nonumber\\
    & 0\leq \lambda_k \leq \lambda_k^{\max}, \forall k\in\mathcal{K}, \nonumber
\end{eqnarray}
where ASE is given in \eqref{eq:ASE_USDMA} and $p_{\rm s}$ is given in \eqref{eq:Ps_USDMA}.

Note that in unbiased $U$-SDMA networks, Problem~$\mathcal{P}_o$ is a linear programming problem \cite{Boyd04}, which can be solved efficiently. To get more insights, we will derive a more explicit solution. From Lemma~\ref{Lemma:Ps_USDMA}, we know that if $\Theta>\max_k p_{\rm s}\left(k\right)$, there is no feasible solution, which implies that whatever the BS density is, the network cannot achieve such a link reliability requirement. Thus, in the following analysis, we assume $\Theta\leq\max_k p_{\rm s}\left(k\right)$.

Denote $y_k=p_{\rm s}\left(k\right)\lambda_k$, then Problem~$\mathcal{P}_0$ is equivalent to
\begin{eqnarray} \label{eq:Problem_eqv}
  \mathcal{P}_{U-\rm{SDMA}} : \underset{\left\{y_k\right\}}{\text{maximize}} & \sum_{k=1}^K y_k \\
  \text{subject to} & \sum_{k=1}^K b_k y_k \geq 0, \nonumber\\
    & 0\leq y_k \leq y_k^{\max}, \forall k\in\mathcal{K}, \nonumber
\end{eqnarray}
where $b_k=\left(\frac{P_k}{U}\right)^\delta \left[1-\frac{\Theta}{p_{\rm s}\left(k\right)}\right]$.
Note that $b_k<0$ if $p_{\rm s}\left(k\right)<\Theta$. To find the solution, we start from $\left\{y_k^\star=y_k^{\max}\right\}$. If $\sum_{j=1}^K b_j y_j^\star \geq 0$, then the optimal solution is to activate all the BSs. On the contrary, if $\sum_{j=1}^K b_j y_j^\star < 0$, it means we need to decrease some $y_k^\star$ until  $\sum_{j=1}^K b_j y_j^\star = 0$. Without loss of generality, assume some of $\left\{b_k\right\}$, i.e., $\left\{b_1,\ldots,b_n\right\}$ where $n\leq K$, are negative, and $b_1\leq b_2 \leq b_n <0$. Then, we need to firstly decrease $y_1^\star$, as $b_1$ has the most negative effect. If $y_1^\star=0$ and $\sum_{j=1}^K b_j y_j^\star < 0$, then $y_2^\star$ should be decreased. So on and so forth until $\sum_{j=1}^K b_j y_j^\star = 0$. The formal solution is described in Algorithm~\ref{alg:alg}.

\begin{algorithm}
\caption{Finding the optimal solution of Problem~$\mathcal{P}_0$ in unbiased $U$-SDMA HetNets}
\label{alg:alg}
    \begin{algorithmic}[1]
        \State Initialize $y_k^{\star} \gets y_k^{\max}$ for $k\in\mathcal{K}$ and $n \gets 1$;
        \While{$\sum_{k=1}^K b_k y_k^{\star} <0$}
            \State $i \gets$ the index of the $n$th minimal value among $\left\{b_k\right\}$;
            \State $y_i^\star \gets 0$;
            \If{$\sum_{k=1}^K b_k y_k^{\star} \geq 0$}
                \State $y_i^\star \gets \frac{1}{-b_i} \sum_{k=1}^K b_k y_k^{\star}$;
                \State $\textbf{break}$;
            \EndIf
            \State $n \gets n+1$;
        \EndWhile
    \end{algorithmic}
\end{algorithm}

From Algorithm~\ref{alg:alg}, we see that the value of $b_k$ is related to the transmit power $P_k$ and the number of antennas $M_k$. A negative $b_k$ will have a negative effect on $p_{\rm s}$. If the BSs in the $k$th tier have a smaller $M_k$ and a larger $P_k$, the value of $b_k$ will be negative and smaller. Therefore, to achieve a high link reliability requirement, the BSs of the tiers with a small $b_k$ should be switched off. Note that one special case is that all the BSs have the same number of antennas, i.e., $M_k=M$ for $\forall k\in\mathcal{K}$. In this special case, $p_{\rm s}\left(k\right)$ is the same for all $k\in\mathcal{K}$. Thus, all the values of $\left\{b_k\right\}$ are either greater than 0 or less than 0, and there is no tradeoff since $p_{\rm s}$ becomes a constant. SISO HetNets belong to this special case.

\subsection{A Demonstration of the Tradeoff in Unbiased $U$-SDMA Networks}

In this subsection, we will demonstrate how to achieve the maximum ASE given the link reliability requirement in a 3-tier HetNet, consisting of micro-BSs, pico-BSs and femto-BSs, where $\left[M_1,M_2,M_3\right]=\left[4,2,2\right]$, $\left[P_1,P_2,P_3\right]=\left[6.3, 0.13, 0.05\right]$ Watts \cite{Auer11}, and the actual BS densities are $\left[\lambda_1^{\max},\lambda_2^{\max},\lambda_3^{\max}\right]=[1,2,5]\times10^{2}$ per ${\rm km}^2$. We consider two cases with $U=1$ and $U=2$, respectively.

\begin{figure}
    \begin{center}
    \scalebox{0.7}{\includegraphics{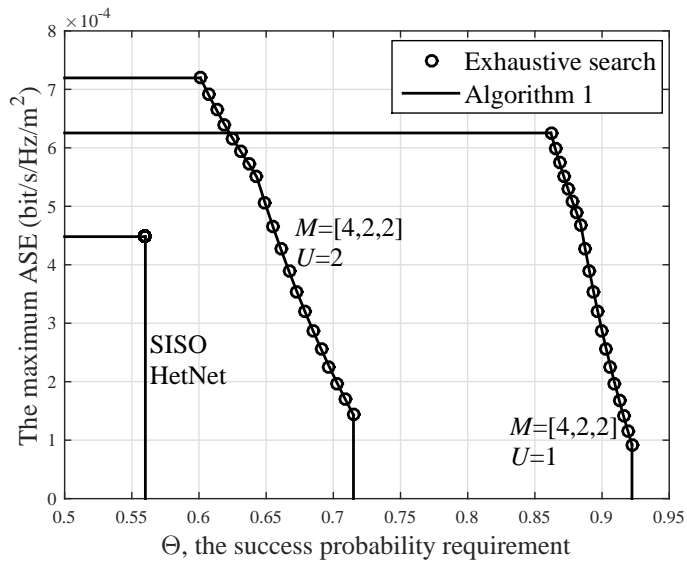}}
    \end{center}
    \caption{The maximum ASE with different requirements of the success probability, with $\alpha=4$, $\hat\gamma=0$ dB, $\left[M_1,M_2,M_3\right]=\left[4, 2, 2\right]$, and $\left[P_1,P_2,P_3\right]=\left[6.3, 0.13, 0.05\right]$ Watts. The actual BS densities are $\left[\lambda_1^{\max},\lambda_2^{\max},\lambda_3^{\max}\right]=\left[1, 2, 5\right]\times10^{2}$ per ${\rm km}^2$.}
    \label{fig:Tradeoff_USDMA}
\end{figure}

In Fig.~\ref{fig:Tradeoff_USDMA}, we show the tradeoff between the maximum ASE and the link reliability requirement $\Theta$. As a benchmark, we also consider the SISO HetNet with the same transmit power $\left\{P_k\right\}$ and actual BS densities $\left\{\lambda_k^{\max}\right\}$. From Fig.~\ref{fig:Tradeoff_USDMA}, we can find that: 1) Compared with the SISO HetNet, deploying multi-antenna BSs can increase both the ASE and link reliability. For example, in the SISO HetNet, $p_{\rm s}$ is always 0.56 whatever the BS density is. But when using multi-antenna BSs, the network can achieve better ASE and $p_{\rm s}$. 2) Comparing the cases with $U=1$ and $U=2$, we see that serving one user at each time slot (i.e., TDMA) can obtain a higher maximum $p_{\rm s}$. This is because the channel gains of the interference links are small when $U$ is small. However, the maximum ASE is higher when $U=2$, compared with $U=1$, since there are more links in a unit area.

\begin{figure}
    \begin{center}
    \scalebox{0.7}{\includegraphics{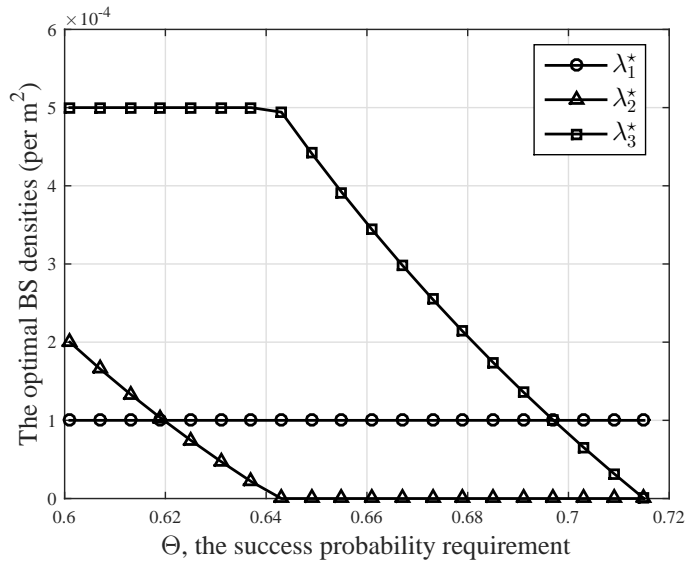}}
    \end{center}
    \caption{The optimal BS densities with different requirements of the success probability, with $\alpha=4$, $\hat\gamma=0$ dB, $\left[M_1,M_2,M_3\right]=\left[4, 2, 2\right]$, $U=2$, and $\left[P_1,P_2,P_3\right]=\left[6.3, 0.13, 0.05\right]$ Watts. The actual BS densities are $\left[\lambda_1^{\max},\lambda_2^{\max},\lambda_3^{\max}\right]=\left[1, 2, 5\right]\times10^{2}$ per ${\rm km}^2$.}
    \label{fig:OptimalBSDensity_USDMA}
\end{figure}

The corresponding optimal BS densities of the case with $U=2$ are provided in Fig.~\ref{fig:OptimalBSDensity_USDMA}. When $\Theta>0.58$, using $\left\{\lambda_k^{\max}\right\}$ cannot satisfy the link reliability constraint. From calculation, we find that the 2nd tier has the minimal value of $\left(\frac{P_k}{U}\right)^\delta \left[1-\frac{\Theta}{p_{\rm s}\left(k\right)}\right]$, which means we need to first decrease the BS density of the 2nd tier, as confirmed in Fig.~\ref{fig:OptimalBSDensity_USDMA}. Comparing the 3rd tier with the 2nd tier, we can find that $M_2=M_3$, but $P_2>P_3$, which means that BSs in the 2nd tier have more negative effects on the link reliability as they will cause higher interference. This is the reason why we need to decrease the BS density of the 2nd tier to improve the link reliability.

\subsection{ASE and Link Reliability Tradeoff in General MIMO HetNets}

Now, we will concentrate on the tradeoff in a general multiuser MIMO HetNet. Since the analysis becomes extremely difficult with the exact expression of $p_{\rm s}$ in \eqref{eq:Psk_Matrix}, we resort to the asymptotic expression in \eqref{eq:Psk_Asy}. By substituting \eqref{eq:Psk_Asy} into \eqref{eq:ASE_def}, the ASE is given by
\begin{equation} \label{eq:ASE_Vector}
    {\rm ASE}\sim\hat{\gamma}^{-\delta}{\rm sinc}\left(\delta\right) \log_{2}\left(1+\hat{\gamma}\right) \frac{\left(\mathbf{c}_{1}^{T}\boldsymbol{\lambda}\right) \left(\mathbf{c}_{2}^{T}\boldsymbol{\lambda}\right)} {\mathbf{d}^{T}\boldsymbol{\lambda}},
\end{equation}
where $\mathbf{c}_{1}\!=\!\left[c_{11},c_{12},\ldots c_{1K}\right]^{T}$, and its element is given as $c_{1i}\!=\!P_{i}^{\delta}\!B_{i}^{\delta}$, while $\mathbf{c}_{2}\!=\!\left[c_{21},c_{22},\ldots c_{2K}\right]^{T}$, in which $c_{2i}=U_{i}^{1-\delta}B_{i}^{-\delta}\frac{\Gamma\left(D_{i}+\delta\right)}{\Gamma\left(D_{i}\right)}$, and $\mathbf{d}$ has the same expression as in \eqref{eq:Ps_Asy_Vector}.

From \eqref{eq:ASE_Vector}, we can find that the ASE is a quadratic-over-linear function with respect to $\boldsymbol{\lambda}$. It means, in general, we cannot guarantee that increasing the BS density will always increase the ASE, and the effect of the BS density on the ASE becomes complicated. In the following, we consider the same problem $\mathcal{P}_o$ as in \eqref{eq:Problem_ori}, to find the optimal BS densities that can achieve the maximum ASE. Based on \eqref{eq:Ps_Asy_Vector} and \eqref{eq:ASE_Vector}, the optimization problem is given as
\begin{eqnarray} \label{eq:Problem_Gen}
    \mathcal{P}_{1} : \underset{\boldsymbol{\lambda}}{\text{maximize}} & \frac{\left(\mathbf{c}_{1}^{T}\boldsymbol{\lambda}\right) \left(\mathbf{c}_{2}^{T}\boldsymbol{\lambda}\right)} {\mathbf{d}^{T}\boldsymbol{\lambda}} \\
    \text{subject to} & \hat{\gamma}^{-\delta}{\rm sinc}\left(\delta\right) \frac{\mathbf{c}^{T}\boldsymbol{\lambda}}{\mathbf{d}^{T}\boldsymbol{\lambda}} \geq \Theta, \nonumber\\
    & 0\leq \lambda_k \leq \lambda_k^{\max}, k\in\mathcal{K}. \nonumber
\end{eqnarray}
Since the maximum $p_{\rm s}$ is obtained by activating only one tier which has the maximal value of $\frac{\Gamma\left(D_{i}+\delta\right)/\Gamma\left(D_{i}\right)} {\Gamma\left(U_{i}+\delta\right)/\Gamma\left(U_{i}\right)}$ (cf. Lemma~\ref{Lemma:MaxPs_Asy}), and $p_{\rm s}^{\max}\sim \hat{\gamma}^{-\delta}{\rm sinc}\left(\delta\right) \max_i \frac{\Gamma\left(D_{i}+\delta\right)/\Gamma\left(D_{i}\right)} {\Gamma\left(U_{i}+\delta\right)/\Gamma\left(U_{i}\right)}$, we only consider the case when $\Theta\leq p_{\rm s}^{\max}$.

Problem $\mathcal{P}_{1}$ is a non-convex problem since the ASE is not a concave function with respect to the BS density. Fortunately, Dinkelbach has proposed an algorithm to solve the nonlinear fractional problems \cite{Dinkelbach67}. By defining $N\left(\boldsymbol{\lambda}\right)= \left(\mathbf{c}_{1}^{T}\boldsymbol{\lambda}\right)\left(\mathbf{c}_{2}^{T}\boldsymbol{\lambda}\right)$ and \begin{equation} \label{eq:Problem_Gen_Func}
    F\left(t\right)=\max_{\boldsymbol{\lambda}}\left\{ N\left(\boldsymbol{\lambda}\right)-t\mathbf{d}^{T}\boldsymbol{\lambda} \mid\boldsymbol{\lambda}\in\mathcal{S}\right\},
\end{equation}
where $\mathcal{S}=\left\{\boldsymbol{\lambda} \mid \hat{\gamma}^{-\delta}{\rm sinc}\left(\delta\right) \frac{\mathbf{c}^{T}\boldsymbol{\lambda}}{\mathbf{d}^{T}\boldsymbol{\lambda}} \geq \Theta, 0\leq \lambda_k \leq \lambda_k^{\max}, k\in\mathcal{K}\right\}$, the optimal BS density $\boldsymbol{\lambda}^{\star}$ can be obtained by finding $t^{\star}$ such that $F\left(t^{\star}\right)=0$ \cite{Dinkelbach67}. Specifically, by iterating $t^{\left(i\right)}=N\left(\boldsymbol{\lambda} ^\star\right)/\left(\mathbf{d}^{T}\boldsymbol{\lambda} ^\star\right)$, where $\boldsymbol{\lambda} ^\star$ is the optimal solution of the right hand side of \eqref{eq:Problem_Gen_Func}, $F\left(t^{\left(i\right)}\right)$ will converge to 0 and $\boldsymbol{\lambda}^\star$ will be the optimal BS density. However, the optimization problem in \eqref{eq:Problem_Gen_Func} is still a non-convex problem due to the non-convex function $N\left(\boldsymbol{\lambda}\right)$. To resolve this difficulty, we resort to the sequential convex programming (SCP) by approximating $N\left(\boldsymbol{\lambda}\right)$ with the first order Taylor expansion, given by $N\left(\boldsymbol{\lambda}\right)\approx N\left(\boldsymbol{\lambda}^{\left(n\right)}\right)+\nabla N\left(\boldsymbol{\lambda}^{\left(n\right)}\right)^{T}\left(\boldsymbol{\lambda}- \boldsymbol{\lambda}^{\left(n\right)}\right)$. It is a simple but effective method and has wide applications, e.g., see \cite{Facchinei14}. Therefore, given the $n$th iterative $\boldsymbol{\lambda}^{\left(n\right)}$ and the $i$th iterative $t^{\left(i\right)}$, the convex optimization problem is given by
\begin{eqnarray} \label{eq:Problem_Iteration}
    \mathcal{P}_{2}\left(\boldsymbol{\lambda}^{\left(n\right)},t^{\left(i\right)}\right) : \underset{\boldsymbol{\lambda}}{\text{maximize}} & N\left(\boldsymbol{\lambda}^{\left(n\right)}\right)+\nabla N\left(\boldsymbol{\lambda}^{\left(n\right)}\right)^{T} \left(\boldsymbol{\lambda}-\boldsymbol{\lambda}^{\left(n\right)}\right) -t^{\left(i\right)}\mathbf{d}^{T}\boldsymbol{\lambda} \nonumber\\
    \text{subject to} & \left(\hat{\gamma}^{-\delta}{\rm sinc}\left(\delta\right)\mathbf{c}^{T} -\Theta\mathbf{d}^{T}\right) \boldsymbol{\lambda}\geq0, \nonumber\\
    & 0\leq \boldsymbol{\lambda} \leq \boldsymbol{\lambda}^{\max}. \nonumber
\end{eqnarray}
Then, the optimal BS density can be obtained by Algorithm~\ref{alg:alg2}. Note that since SCP can only obtain a local maximum, we will randomly generate multiple initial values of  $\boldsymbol{\lambda}^{\left(0\right)}$ to find a better solution.


\begin{algorithm}
\caption{The Locally Optimal BS Densities in General MIMO HetNets}
\label{alg:alg2}
    \begin{algorithmic}[1]
        \State  Initialize $\boldsymbol{\lambda}^{\left(0\right)} \gets$ random value between $\left[0,\boldsymbol{\lambda}^{\max}\right]$, $n \gets 0$, $i \gets 0$ and assign $\varepsilon$ a small value;
        \State $t^{\left(i\right)} \gets \frac{\left(\mathbf{c}_{1}^{T}\boldsymbol{\lambda}^{\left(n\right)}\right) \left(\mathbf{c}_{2}^{T}\boldsymbol{\lambda}^{\left(n\right)}\right)} {\mathbf{d}^{T}\boldsymbol{\lambda}^{\left(n\right)}}$; \label{algstep:t}
        \State Solve Problem $\mathcal{P}_{2}\left(\boldsymbol{\lambda}^{\left(n\right)},t^{\left(i\right)}\right)$ and obtain the optimal value $\boldsymbol{\lambda}^{\star}$; \label{algstep:solve}
        \If{$\left\Vert \boldsymbol{\lambda}^{\star}-\boldsymbol{\lambda}^{\left(n\right)}\right\Vert /\left\Vert \boldsymbol{\lambda}^{\left(n\right)}\right\Vert \geq\varepsilon$}
            \State $n\gets n+1$, $\boldsymbol{\lambda}^{\left(n\right)} \gets \boldsymbol{\lambda}^{\star}$, and go to Step~\ref{algstep:solve};
        \Else
            \If{$\left(\mathbf{c}_{1}^{T}\boldsymbol{\lambda}^{\star}\right) \left(\mathbf{c}_{2}^{T}\boldsymbol{\lambda}^{\star}\right) -t^{\left(i\right)}\mathbf{d}^{T}\boldsymbol{\lambda}^{\star}\geq\varepsilon$}
                \State $i \gets i+1$, and go to Step~\ref{algstep:t};
            \Else
                \State \textbf{return} $\boldsymbol{\lambda}^{\star}$;
            \EndIf
        \EndIf
    \end{algorithmic}
\end{algorithm}


\subsection{A Demonstration of the Tradeoff in General MIMO HetNets}

\begin{figure}
    \begin{center}
    \scalebox{0.7}{\includegraphics{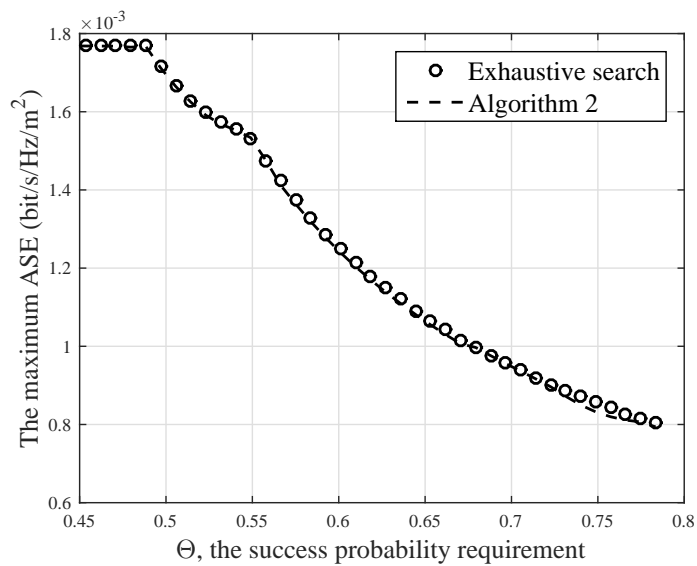}}
    \end{center}
    \caption{The maximum ASE with different requirements of the success probability, with $\alpha=4$, $\hat\gamma=5$ dB, $\left[M_1,M_2,M_3\right]=\left[8, 4, 1\right]$, $\left[U_1,U_2,U_3\right]=\left[4, 1, 1\right]$, $B_k=1/U_k$ for $k=1,2,3$, and $\left[P_1,P_2,P_3\right]=\left[6.3, 0.13, 0.05\right]$ Watts. The actual BS densities are $\boldsymbol{\lambda}^{\max}=\left[1, 5, 10\right]\times10^{2}$ per ${\rm km}^2$.}
    \label{fig:Tradeoff_General}
\end{figure}

\begin{figure}
    \begin{center}
    \scalebox{0.7}{\includegraphics{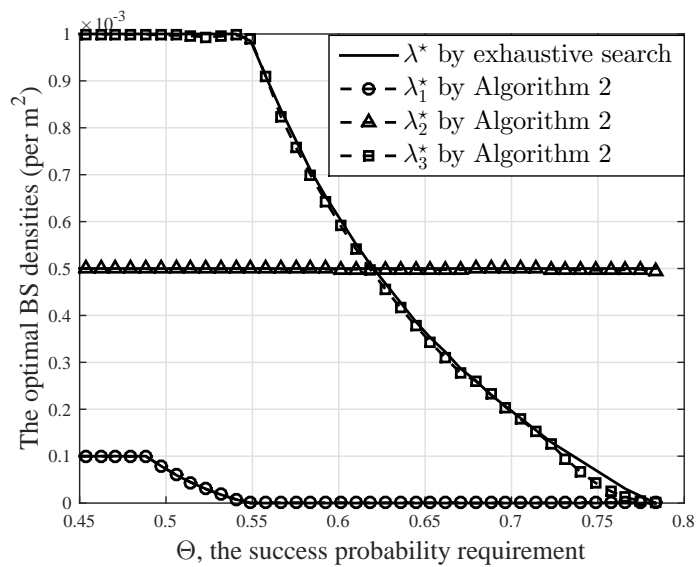}}
    \end{center}
    \caption{The optimal BS densities with different requirements of the success probability, with $\alpha=4$, $\hat\gamma=5$ dB, $\left[M_1,M_2,M_3\right]=\left[8, 4, 1\right]$, $\left[U_1,U_2,U_3\right]=\left[4, 1, 1\right]$, $B_k=1/U_k$ for $k=1,2,3$, and $\left[P_1,P_2,P_3\right]=\left[6.3, 0.13, 0.05\right]$ Watts. The actual BS densities are $\boldsymbol{\lambda}^{\max}=\left[1, 5, 10\right]\times10^{2}$ per ${\rm km}^2$.}
    \label{fig:OptimalBSDensity_General}
\end{figure}

In this subsection, we will use Algorithm~\ref{alg:alg2} to evaluate the tradeoff between the ASE and the link reliability. We consider a 3-tier HetNet, where $\left[M_1,M_2,M_3\right]=\left[8,4,1\right]$, $\left[U_1,U_2,U_3\right]=\left[4,1,1\right]$, $B_k=1/U_k$ for $k=1,2,3$,  $\left[P_1,P_2,P_3\right]=\left[6.3, 0.13, 0.05\right]$ Watts, and the actual BS densities are $\boldsymbol{\lambda}^{\max}=\left[\lambda_1^{\max},\lambda_2^{\max},\lambda_3^{\max}\right] =[1,5,10]\times10^{2}$ per ${\rm km}^2$. For Algorithm~\ref{alg:alg2}, we generate 20 randomly initial values of $\boldsymbol{\lambda}^{\left(0\right)}$, and set $\varepsilon = 10^{-6}$. In Fig.~\ref{fig:Tradeoff_General}, we show the tradeoff between the maximum ASE and the link reliability requirement $\Theta$, while the corresponding optimal BS densities are shown in Fig.~\ref{fig:OptimalBSDensity_General}.

First, we find from both figures that the proposed algorithm can achieve almost the same performance as using the exhaustive search, while the proposed algorithm runs much faster. Second, from Fig.~\ref{fig:Tradeoff_General}, we find that in general cases, there exists a tradeoff between the ASE and link reliability, and the higher the link reliability requires, the lower the ASE can be achieved. More interestingly, from Fig.~\ref{fig:OptimalBSDensity_General}, we find that: 1) Even in general HetNets, the maximal ASE is achieved by activating all the BSs. 2) With the increasing requirement of the link reliability, the BS density will decrease from one tier to another tier. Moreover, only when the BS density decrease to 0, the BS density from another tier will start to decrease, which is the same as $U$-SDMA HetNets. Thus, we can infer that for a given $\Theta$, different tiers have different influences on the network, and there is an ordering of such influences. Recall that in $U$-SDMA networks, the ordering can be explicitly obtained from $\left\{b_k\right\}$, where $b_{k}=\left(\frac{P_{k}}{U}\right)^{\delta}\left[1-\frac{\Theta}{p_{\rm s}\left(k\right)}\right]$, and the tier with the minimal negative $b_k$ among $k\in\mathcal{K}$ has the most negative effect on the link reliability. However, in the general MIMO HetNets, so far we are unable to derive an exact expression to find such an ordering, since both signal and interference distributions are complicated. But with our algorithm, we can numerically find the effects of different tiers on the link reliability. For example, from Fig.~\ref{fig:OptimalBSDensity_General}, we can see that tier 1 (circle points) has the most negative effect on the link reliability. When $\Theta\gtrsim0.5$, the BSs from tier 1 need to be deactivated to guarantee the link reliability. It is because the interference caused by tier 1 is large due to the high multiplexing gain $U_1=4$.

\section{Conclusions} \label{Sec:Conclusions}

In this paper, we developed a new set of analytical results for performance analysis of downlink multiuser MIMO HetNets, where multi-antenna BSs use SDMA (ZF precoding) to serve multiple users. Both exact and asymptotic expressions of the success probability were provided. We focused on the effect of the BS densities, and proved that there is no invariance property for the success probability in general MIMO HetNets. A unique tradeoff between the link reliability and the ASE was revealed. By using the proposed algorithms, we found the optimal BS densities to achieve the maximum ASE while guaranteeing a given link reliability requirement. It was shown that the maximum ASE of the network is achieved by activating all the BSs, while the maximum link reliability is achieved by activating only one tier of BSs.

The link reliability vs. ASE tradeoff analyzed in this paper provides a new perspective in designing multiuser MIMO HetNets, which is fundamentally different from SISO HetNets. Both the link reliability and ASE should be considered when evaluating different transmission techniques for HetNets. A basic network model is considered in this paper, while more general models require further investigation, such as multi-slope path loss models \cite{Zhang15}, together with more sophisticated techniques such as interference coordination \cite{Li15} and load balancing \cite{Gupta14}.

\appendix

\subsection{Proof of Theorem~\ref{Thm:Psk_General}} \label{App:Psk_General}


The success probability in \eqref{eq:Psk_ori} can be rewritten as
\begin{equation}
    p_{{\rm s}}\left(k\right)= \mathbb{P}\left(g_{x_0,k}\geq r_{k}^{\alpha}\hat{\gamma}\sum_{j=1}^{K}\sum_{x\in\Psi_{j}\backslash\left\{ x_{0}\right\} }\frac{P_{j}U_{k}}{P_{k}U_{j}}g_{x,j}\left\Vert x\right\Vert ^{-\alpha}\right).
\end{equation}
Since $g_{x_0,k}\stackrel{d}{\sim}{\rm Gamma}\left(D_k,1\right)$, the above equality can be expressed as
\begin{equation}
    p_{{\rm s}}\left(k\right)=\mathbb{E}_{s}\left[\sum_{n=0}^{D_{k}-1} \mathbb{E}_{I}\left[\frac{\left(sI\right)^{n}}{n!}e^{-sI}\right]\right],
\end{equation}
where $s\triangleq\hat\gamma r_k^\alpha$ and $I\triangleq\sum_{j=1}^{K}\sum_{x\in\Psi_{j}\backslash\left\{ x_{0}\right\} }\frac{P_{j}U_{k}}{P_{k}U_{j}}g_{x,j}\left\Vert x\right\Vert ^{-\alpha}$. Note that for a fixed $s$, $\mathbb{E}_I\left[e^{-sI}\right]$ is the Laplace
transform of $I$, denoted as $\mathcal{L}_I\left(s\right)$. Following the property of the Laplace transform, we have $\mathbb{E}_{I}\left[I^{n}e^{-sI}\right]= \left(-1\right)^{n}\mathcal{L}_{I}^{\left(n\right)}\left(s\right)$, where $\mathcal{L}_{I}^{\left(n\right)}\left(s\right)$ is the $n$th derivative of $\mathcal{L}_{I}\left(s\right)$. Then, we have
\begin{equation} \label{eq:Psk_LaplaceDef}
    p_{{\rm s}}\left(k\right)=\mathbb{E}_{s}\left[\sum_{n=0}^{D_{k}-1} \frac{\left(-s\right)^{n}}{n!}\mathcal{L}_{I}^{\left(n\right)}\left(s\right)\right].
\end{equation}



To derive $p_{{\rm s}}\left(k\right)$ based on \eqref{eq:Psk_LaplaceDef}, we start from the Laplace transform of $I$ for a fixed $s$, given by
\begin{equation}
    \mathcal{L}_{I}\left(s\right)= \mathbb{E}\left[\exp\left(-s\sum_{j=1}^{K}\sum_{x\in\Psi_{j}\backslash\left\{ x_{0}\right\} }\frac{P_{j}U_{k}}{P_{k}U_{j}}g_{x,j}\left\Vert x\right\Vert ^{-\alpha}\right)\middle|s\right].
\end{equation}
Since $\left\{\Psi_j:j\in\mathcal{K}\right\}$ and $\left\{g_{x,j}:x\in\Psi_b, j\in\mathcal{K}\right\}$ are independent, the above equality can be written as
\begin{eqnarray}
  \mathcal{L}_{I}\left(s\right)  &=&  \prod_{j=1}^{K}\mathbb{E}_{\Psi_{j}}  \left[\prod_{x\in\Psi_{j}\backslash\left\{ x_{0}\right\} }\mathbb{E}_{g_{x}}\left[\exp\left(-s\frac{P_{j}U_{k}}{P_{k}U_{j}}g_{x,j}\left\Vert x\right\Vert ^{-\alpha}\right)\right]\right] \nonumber\\
  & \stackrel{\left({\rm a}\right)}{=}&  \prod_{j=1}^{K}\mathbb{E}_{\Psi_{j}}  \left[\prod_{x\in\Psi_{j}\backslash\left\{ x_{0}\right\} }\left(1+s\frac{P_{j}U_{k}}{P_{k}U_{j}}\left\Vert x\right\Vert ^{-\alpha}\right)^{-U_{j}}\right],
\end{eqnarray}
where (a) follows from $g_{x,j}\stackrel{d}{\sim}{\rm Gamma}\left(U_j,1\right)$ for $x\in\Psi_j$. Then, the Laplace transform of $I$ can be derived using the probability generating functional (PGFL) \cite{Haenggi12}, which is given as
\begin{equation} \label{eq:Laplace_ori}
    \mathcal{L}_{I}\left(s\right)= \exp\left\{ -\pi\sum_{j=1}^{K} \lambda_{j}  \int_{r_{j}^{2}}^{\infty}  \left[1-\left(1+s\frac{P_{j}U_{k}}{P_{k}U_{j}}v^{-\frac{\alpha}{2}}\right)^{-U_{j}} \right]dv\right\},
\end{equation}
where $r_j$ is the minimal distance between the typical user and the BSs of the $j$th type. Based on \eqref{eq:Laplace_ori}, the $n$th derivative of $\mathcal{L}_I\left(s\right)$ with respect to $s$ can be written as the following recursive form
\begin{equation} \label{eq:Laplace_div}
    \mathcal{L}_{I}^{\left(n\right)}\left(s\right)=\sum_{i=0}^{n-1}\left(\begin{array}{c}
    n-1\\
    i
    \end{array}\right) \left[\pi\sum_{j=1}^{K}\lambda_{j}\int_{r_{j}^{2}}^{\infty} \left(-1\right)^{n-i}\left(U_{j}\right)_{n-i} \frac{\left(\frac{P_{j}U_{k}}{P_{k}U_{j}}v^{-\frac{\alpha}{2}}\right)^{n-i}dv} {\left(1+s\frac{P_{j}U_{k}}{P_{k}U_{j}}v^{-\frac{\alpha}{2}}\right)^{U_{j}+n-i}}\right] \mathcal{L}_{I}^{\left(i\right)}\left(s\right),
\end{equation}
where $\left(U_{j}\right)_{n-i}$ is the Pochhammer symbol.

Denote $x_n = \frac{1}{n!}\left(-s\right)^n \mathcal{L}_{I}^{\left(n\right)}\left(s\right)$ for $n\geq0$, then it shows that $p_{{\rm s}}\left(k\right)=\mathbb{E}_{s}\left[\sum_{n=0}^{D_{k}-1}x_{n}\right]$. The expression of $x_0$ can be obtained directly from \eqref{eq:Laplace_ori}, while substituting \eqref{eq:Laplace_div} to $x_n$, we can get for $n\geq1$,
\begin{equation}
    x_{n}=\pi\sum_{i=0}^{n-1}x_{i}\frac{n-i}{n} \left[\sum_{j=1}^{K}\lambda_{j}\int_{r_{j}^{2}}^{\infty} \frac{\left(U_{j}\right)_{n-i}}{\left(n-i\right)!} \frac{\left(s\frac{P_{j}U_{k}}{P_{k}U_{j}}v^{-\frac{\alpha}{2}}\right)^{n-i}dv} {\left(1+s\frac{P_{j}U_{k}}{P_{k}U_{j}}v^{-\frac{\alpha}{2}}\right)^{U_{j}+n-i}}\right].
\end{equation}

In the expression of $x_n$, the integration limits are from $r_j^2$ to infinity. As the typical user is associated with the BS in the $k$th tier, it implies that $P_k B_k r_k^{-\alpha} \geq P_j B_j r_j^{-\alpha}$ for $j\in\mathcal{K}$, which is equivalent to $r_j^2\geq \left(\frac{P_jB_j}{P_kB_k}\right)^\delta r_k^2$, where $\delta\triangleq\frac{2}{\alpha}$. Moreover, substituting $s=\hat\gamma r_k^\alpha$ to $x_0$, the expression of $x_0$ is given by
\begin{equation}
    x_{0}=\exp\left\{ -\pi\sum_{j=1}^{K}\lambda_{j} \int_{\left(\frac{P_{j}B_{j}}{P_{k}B_{k}}\right)^{\delta}r_{k}^{2}}^{\infty} \left[1-\left(1+\hat{\gamma}r_{k}^{\alpha} \frac{P_{j}U_{k}}{P_{k}U_{j}}v^{-\frac{\alpha}{2}}\right)^{-U_{j}}\right]dv\right\},
\end{equation}
and for $n\geq1$,
\begin{equation}
    x_{n}= \pi\sum_{i=0}^{n-1}x_{i}\frac{n-i}{n} \left[\sum_{j=1}^{K}\lambda_{j}  \int_{\left(\frac{P_{j}B_{j}}{P_{k}B_{k}}\right)^{\delta}r_{k}^{2}}^{\infty}  \frac{\left(U_{j}\right)_{n-i}}{\left(n-i\right)!}
    \frac{\left(\hat{\gamma}r_{k}^{\alpha} \frac{P_{j}U_{k}}{P_{k}U_{j}}v^{-\frac{\alpha}{2}}\right)^{n-i}} {\left(1+\hat{\gamma}r_{k}^{\alpha}\!\frac{P_{j}U_{k}}{P_{k}U_{j}} v^{-\frac{\alpha}{2}}\right)^{U_{j}+n-i}}dv\right].
\end{equation}
Denote $q_i$ as \eqref{eq:qi}, then with some manipulation, the expression of $x_n$ can be simplified as
\begin{equation} \label{eq:x0}
    x_{0}=\exp\left[-\frac{\pi r_{k}^{2}}{P_{k}^{\delta}B_{k}^{\delta}}\left(q_{0}-A\right)\right],
\end{equation}
and for $n\geq1$,
\begin{equation} \label{eq:Recurrent}
    x_n = -\frac{\pi r_{k}^{2}}{P_{k}^{\delta}B_{k}^{\delta}} \sum_{i=0}^{n-1}\frac{n-i}{n}q_{n-i}x_i.
\end{equation}

To solve the recurrent relation of $x_n$, we define two power series $F\left(z\right)$ and $G\left(z\right)$ as
\begin{equation}
    F\left(z\right)\triangleq\sum_{n=0}^{\infty}q_{n}z^{n},\quad G\left(z\right)\triangleq\sum_{n=0}^{\infty}x_{n}z^{n}.
\end{equation}
Using the properties of $F^{'}\left(z\right)=\sum_{n=0}^{\infty}nq_{n}z^{n-1}$ and $F\left(z\right)G\left(z\right)=\sum_{n=0}^{\infty}\left(\sum_{i=0}^{n}q_{n-i}x_{i}\right)z^{n}$, from \eqref{eq:Recurrent}, we can get the equality as
\begin{equation}
    zG^{'}\left(z\right) = -\frac{\pi r_{k}^{2}}{P_{k}^{\delta}B_{k}^{\delta}} \left(zF^{'}\left(z\right)G\left(z\right)\right).
\end{equation}
Then, by solving the above differential equation, we have $G\left(z\right)=c\exp\left(-\frac{\pi r_{k}^{2}}{P_{k}^{\delta}B_{k}^{\delta}} F\left(z\right)\right)$, where $c$ is a constant to be determined. Since $G\left(0\right)=x_0$ and $G\left(0\right)=c\exp\left(-\frac{\pi r_{k}^{2}}{P_{k}^{\delta}B_{k}^{\delta}}q_0\right)$, from \eqref{eq:x0}, we get $c=\exp\left(\frac{\pi r_k^2}{P_k^\delta B_k^\delta} A\right)$. Thus, $G\left(z\right)$ is given by
%
%
%
%
\begin{equation}
    G\left(z\right)=\exp\left[-\frac{\pi r_{k}^{2}}{P_{k}^{\delta}B_{k}^{\delta}}\left(F\left(z\right)-A\right)\right].
\end{equation}

Recalling that $p_{\rm s}\left(k\right)=\mathbb{E}_{r_k}\left[\sum_{n=0}^{D_k-1}x_n\right]$, we define
\begin{equation}
    T\left(z\right)\triangleq\mathbb{E}_{r_k}\left[G\left(z\right)\right]=\sum_{n=0}^\infty t_n z^n.
\end{equation}
From \cite[Lemma~3]{Jo12}, the probability density function of $r_k$ is given by
\begin{equation}
    f_{r_{k}}\left(r\right)=2\pi r\left[\sum_{j=1}^{K}\lambda_{j}\left(\frac{P_{j}B_{j}}{P_{k}B_{k}}\right)^{\delta}\right]e^{-\pi r^{2}\left[\sum_{j=1}^{K}\lambda_{j}\left(\frac{P_{j}B_{j}}{P_{k}B_{k}}\right)^{\delta}\right]}.
\end{equation}
Then we can get
\begin{equation}
    T\left(z\right)=\mathbb{E}_{r_{k}}\left[\exp\left[-\frac{\pi r_{k}^{2}}{P_{k}^{\delta}B_{k}^{\delta}} \left(F\left(z\right)-A\right)\right]\right]=\frac{A}{F\left(z\right)}.
\end{equation}
Therefore, the success probability is given by
\begin{equation} \label{eq:psk_summation_tn}
    p_{{\rm s}}\left(k\right) =\sum_{n=0}^{D_{k}-1}t_{n} =\sum_{n=0}^{D_{k}-1}\frac{1}{n!}T^{\left(n\right)}\left(z\right)\mid_{z=0} =A\sum_{n=0}^{D_{k}-1}\frac{1}{n!}\frac{d^{n}}{dz^{n}} \left.\left(\frac{1}{F\left(z\right)}\right)\right|_{z=0}.
\end{equation}

Finally, from \cite[p.~14]{Henrici88}, it can be shown that the first $D_k$ coefficients of the power series $\frac{1}{F\left(z\right)}$ is the first column of the matrix $\mathbf{Q}_{D_k}^{-1}$, where $\mathbf{Q}_{D_k}$ is given in \eqref{eq:Q}. Thus, \eqref{eq:psk_summation_tn} is equivalent to \eqref{eq:Psk_Matrix}, which completes the proof.

\subsection{Proof of Theorem~\ref{Thm:Psk_Asy}} \label{App:Psk_Asy}

By substituting \eqref{eq:qi_Asy} into the power series $F\left(z\right)$, we can obtain
\begin{equation}
    F\left(z\right)=\sum_{i=0}^{\infty}q_{i}z^{i}\sim \sum_{j=1}^{K}\lambda_{j}P_{j}^{\delta}B_{j}^{\delta} \left(\frac{U_{k}B_{k}}{U_{j}B_{j}}\hat{\gamma}\right)^{\delta} \frac{\Gamma\left(U_{j}+\delta\right)}{\Gamma\left(U_{j}\right)}\sum_{i=0}^{\infty} \frac{\delta}{\delta-i}\frac{\Gamma\left(i+1-\delta\right)}{\Gamma\left(i+1\right)}z^{i},
\end{equation}
and it can be expressed as
\begin{equation}
    F\left(z\right)\sim\sum_{j=1}^{K}\lambda_{j}P_{j}^{\delta}B_{j}^{\delta} \left(\frac{U_{k}B_{k}}{U_{j}B_{j}}\hat{\gamma}\right)^{\delta}\frac{\Gamma\left(U_{j}+\delta\right)} {\Gamma\left(U_{j}\right)}\Gamma\left(1-\delta\right)\left(1-z\right)^{\delta}.
\end{equation}
Thus, the power series $T\left(z\right)$ is given by
\begin{equation}
    T\left(z\right)=\frac{A}{F\left(z\right)}\sim \frac{A\frac{1}{\Gamma\left(1-\delta\right)} \left(1-z\right)^{-\delta}} {\sum_{j=1}^{K}\lambda_{j}P_{j}^{\delta}B_{j}^{\delta} \left(\frac{U_{k}B_{k}}{U_{j}B_{j}}\hat{\gamma}\right)^{\delta} \frac{\Gamma\left(U_{j}+\delta\right)}{\Gamma\left(U_{j}\right)}}.
\end{equation}
Based on the above expression, the coefficient $t_n$ is given by
\begin{equation}
    t_{n}=\frac{1}{n!}T^{\left(n\right)}\left(z\right)\mid_{z=0}\sim \frac{A\frac{1}{n!}\left(\delta\right)_{n}\frac{1}{\Gamma\left(1-\delta\right)}} {\sum_{j=1}^{K}\lambda_{j}P_{j}^{\delta}B_{j}^{\delta}\left(\frac{U_{k}B_{k}}{U_{j}B_{j}} \hat{\gamma}\right)^{\delta}\frac{\Gamma\left(U_{j}+\delta\right)}{\Gamma\left(U_{j}\right)}}.
\end{equation}
As $p_{{\rm s}}\left(k\right)=\sum_{n=0}^{D_{k}-1}t_{n}$, and the summation $\sum_{n=0}^{D_{k}-1}\frac{1}{n!}\left(\delta\right)_{n}= \frac{\Gamma\left(D_{k}+\delta\right)}{\Gamma\left(1+\delta\right)\Gamma\left(D_{k}\right)}$, the success probability is given by
\begin{equation}
p_{{\rm s}}\left(k\right)\sim \frac{A\frac{\Gamma\left(D_{k}+\delta\right)} {\Gamma\left(1-\delta\right)\Gamma\left(1+\delta\right)\Gamma\left(D_{k}\right)}} {\sum_{j=1}^{K}\lambda_{j}P_{j}^{\delta}B_{j}^{\delta}\left(\frac{U_{k}B_{k}}{U_{j}B_{j}} \hat{\gamma}\right)^{\delta} \frac{\Gamma\left(U_{j}+\delta\right)}{\Gamma\left(U_{j}\right)}}.
\end{equation}
Using the equality $\Gamma\left(1+\delta\right)\Gamma\left(1-\delta\right)=\frac{\pi\delta}{\sin\left(\pi\delta\right)} =\frac{1}{{\rm sinc}\left(\delta\right)}$, we can obtain \eqref{eq:Psk_Asy}.

\subsection{Proof of Lemma~\ref{Lemma:Ps_Monotonic}} \label{App:Ps_Monotonic}

We consider the function $y=\frac{\mathbf{c}^{T}\boldsymbol{\lambda}}{\mathbf{d}^{T}\boldsymbol{\lambda}}$, where $\boldsymbol{\lambda}=\left[\lambda_{1},\lambda_{2},\ldots\lambda_{K}\right]^{T}$, $\mathbf{c}=\left[c_1,c_2,\ldots,c_K\right]$ and $\mathbf{d}=\left[d_1,d_2,\ldots,d_K\right]$. The partial derivative of $f\left(\boldsymbol{\lambda}\right)$ with respect to $\lambda_i$ is then given by
\begin{equation} \label{eq:PD_Ps}
    \frac{\partial y}{\partial\lambda_{i}} = \frac{\sum_{j=1}^{K}d_{i}d_{j}\lambda_{j}\left(\frac{c_{i}}{d_{i}}-\frac{c_{j}}{d_{j}}\right)} {\left(\mathbf{d}^{T}\boldsymbol{\lambda}\right)^{2}}.
\end{equation}
It shows that changing $\lambda_i$ will not change the sign of $\frac{\partial y}{\partial\lambda_{i}}$, i.e., $y$ is monotonic with respect to $\lambda_i$. Moreover, $y$ is independent of $\boldsymbol{\lambda}$ if $\frac{c_i}{d_i}=\frac{c_j}{d_j}$ for all $i,j\in\left\{ 1,2,\ldots,K\right\}$.

\subsection{Proof of Lemma~\ref{Lemma:MaxPs_Asy}} \label{App:MaxPs}

In this proof, we consider a more general case where the optimization problem is given by
\begin{eqnarray}  
    \mathcal{P}_{w/\:cons} : \underset{\boldsymbol{\lambda}}{\text{maximize}}  & \frac{\mathbf{c}^{T}\boldsymbol{\lambda}}{\mathbf{d}^{T}\boldsymbol{\lambda}} \\
    \text{subject to} & \mathbf{A}\boldsymbol{\lambda}\leq\mathbf{b}, \nonumber\\
    & \boldsymbol{\lambda}\geq\mathbf{0}, \nonumber
\end{eqnarray}
where $\mathbf{A}\boldsymbol{\lambda}\leq\mathbf{b}$ represents arbitrary constraints, $\mathbf{A}$ is a $n\times K$ matrix, and $\mathbf{b}$ is a $n\times1$ vector with positive elements. We will find the optimal solution $\boldsymbol{\lambda}^\star$ in the following derivation.

First we consider the optimization problem without $\mathbf{A}\boldsymbol{\lambda}\leq\mathbf{b}$ constraint, i.e., the problem is given by
\begin{eqnarray} 
    \mathcal{P}_{w/o\:cons} :  \underset{\boldsymbol{\lambda}}{\text{maximize}} & \frac{\mathbf{c}^{T}\boldsymbol{\lambda}}{\mathbf{d}^{T}\boldsymbol{\lambda}} \\
    \text{subject to} & \boldsymbol{\lambda}\geq\mathbf{0}. \nonumber
\end{eqnarray}
Without loss of generality, we assume $\frac{c_{1}}{d_{1}}\geq\frac{c_{2}}{d_{2}}\geq\cdots\geq\frac{c_{K}}{d_{K}}$, then from \eqref{eq:PD_Ps}, we can find that to maximize $p_{\rm s}$, $\lambda_K$ should be 0 since $\frac{\partial p_{{\rm s}}}{\partial\lambda_{K}}\leq0$.

Then, repeating the same procedure, we can find $\lambda_{K-1}=0$, $\lambda_{K-2}=0$, and $\lambda_{2}=0$ successively. Finally, the objective function $p_{\rm s}$ is equal to $\frac{c_1}{d_1}$ with any $\lambda_1>0$, which is the solution of $\mathcal{P}_{w/o\:cons}$.

Secondly, we consider the optimization Problem $\mathcal{P}_{w/\:cons}$, and we want to prove that the optimal $\boldsymbol{\lambda}^\star$ of Problem $\mathcal{P}_{w/o\:cons}$ is also the optimal solution of problem $\mathcal{P}_{w/\:cons}$. To do so, we need to prove 1) $\boldsymbol{\lambda}^\star$ is feasible for $\mathcal{P}_{w/\:cons}$, and 2) the optimal solution of $\mathcal{P}_{w/\:cons}$ is $\boldsymbol{\lambda}^\star$.

To prove the feasibility, we assume $\boldsymbol{\lambda}_0^\star$ and $\boldsymbol{\lambda}_1^\star$ are optimal solutions of $\mathcal{P}_{w/o\:cons}$, but $\mathbf{A}\boldsymbol{\lambda}_1^\star > \mathbf{b}$. Since $\mathbf{b}>0$, and $\boldsymbol{\lambda}_0^\star=k\boldsymbol{\lambda}_1^\star$ for any $k>0$. Then $\exists k>0$, so that $\mathbf{A}\boldsymbol{\lambda}_0^\star=k\mathbf{A}\boldsymbol{\lambda}_1^\star\leq\mathbf{b}$, i.e., there exists an optimal solution of $\mathcal{P}_{w/o\:cons}$, which is feasible to $\mathcal{P}_{w/\:cons}$.

Finally, since $\mathcal{P}_{w/\:cons}$ has one more constraint than $\mathcal{P}_{w/o\:cons}$, the solution of $\mathcal{P}_{w/\:cons}$ should be the subset of the solution of $\mathcal{P}_{w/o\:cons}$ and the maximum value of $\mathcal{P}_{w/\:cons}$ will no greater than the maximum value of $\mathcal{P}_{w/o\:cons}$. As we have proved that the optimal solution of $\mathcal{P}_{w/o\:cons}$ is feasible in $\mathcal{P}_{w/\:cons}$, we can obtain that the maximum value of $\mathcal{P}_{w/\:cons}$ is equal to $\frac{c_1}{d_1}$.

Now we come back to the general MIMO HetNet case. The original problem \eqref{eq:OptiPs_Asy} indicates $\mathbf{A}=\mathbf{I}_K$ and $\frac{c_{i}}{d_{i}}= \frac{\Gamma\left(D_{i}+\delta\right)/\Gamma\left(D_{i}\right)} {\Gamma\left(U_{i}+\delta\right)/\Gamma\left(U_{i}\right)}$. Therefore, the maximum $p_{\rm s}$ is obtained by only deploying one tier which has the maximal value of $\frac{c_{i}}{d_{i}}$.

\bibliographystyle{IEEEtran}
\bibliography{IEEEabrv,Ref}

\end{document}